\documentclass[letter,english]{article}
\usepackage{emulateapj,onecolfloat,epsfig}

\begin{document}
\newcommand {\ds}{\displaystyle} \newcommand
{\sss}{\scriptscriptstyle} \def\sun{\hbox{$\odot$}}

\twocolumn [ \title{Constraints on the Inner Cluster Mass Profile and
the  Power Spectrum Normalization from Strong Lensing Statistics}

\author{Dragan Huterer\footnote{Department of Physics, Case Western
Reserve University, Cleveland, OH~~44106\ }\  and Chung-Pei
Ma\footnote{ Department of Astronomy, University of California,
Berkeley, CA~~94720}}

\begin{abstract}

Strong gravitational lensing is a useful probe of both the intrinsic
properties of the lenses and the cosmological parameters of the
universe.  The large number of model parameters and small sample of
observed lens systems, however, have made it difficult to obtain
useful constraints on more than a few parameters from lensing
statistics. Here we examine how the recent WMAP measurements help
improve the constraining power of statistics from the radio lensing
survey JVAS/CLASS.  We find that the absence of $\theta>3''$ lenses in
CLASS places an upper bound of $\beta<1.25\, (1.60)$ at 68\% (95\%) CL
on the inner density profile, $\rho \propto r^{-\beta}$, of
cluster-sized halos.  Furthermore, the favored power spectrum
normalization is $\sigma_8\ga 0.7$ (95\% CL).  We discuss two
possibilities for stronger future constraints: a positive detection of
at least one large-separation system, and next-generation radio
surveys such as LOFAR.

\end{abstract}
 ]%

\section{Introduction}

Strong gravitational lensing is a sensitive probe of the intrinsic
properties of galaxies, clusters, and dark matter halos as well as of
the underlying cosmological model of the universe.  For instance, some
of the earliest constraints on the cosmological constant were obtained
from statistics of gravitational lensing (e.g. Turner 1990; Kochanek
1996).  The uncertainties in the properties of the lenses themselves,
however, have led some to suggest that lensing studies are better used
as probes of lens properties rather than cosmological parameters
(e.g. Cheng \& Krauss 2000).  This view is particularly pertinent in
light of the recent measurements from the Wilkinson Microwave
Anisotropy Probe (WMAP) and large-scale structure surveys, which, when
combined, can now constrain some of the basic cosmological parameters
to 5\% or better (Spergel et al.\ 2003 and references therein).

Indeed, recent lensing studies have yielded useful constraints on the
density profiles and velocity dispersions of galaxies (Takahashi \&
Chiba 2001; Rusin \& Ma 2001; Oguri, Taruya \& Suto 2001; Keeton \&
Madau 2001; Davis, Huterer \& Krauss 2002; Chae 2002).  Most of these
analyses as well as direct observations (Cohn et al.\ 2001; Treu \&
Koopmans 2002; Winn, Rusin \& Kochanek 2002; Rusin, Kochanek \& Keeton
2003) indicate that the inner density profiles of elliptical galaxies
are close to singular isothermal (SIS): $\rho\propto r^{-\beta}$ with
$\beta \approx 2$.  These analyses have been greatly facilitated by
the completion of the Cosmic Lens All-sky Survey (CLASS; Myers et al.\
2003, Browne et al.\ 2003), which extended the earlier Jodrell Bank-VLA 
Astrometric Survey (JVAS; King et al.\ 1999).  JVAS/CLASS
is the biggest survey with a homogeneous sample of lenses, with a
total of about 16,000 sources and 22 confirmed lensing events. Among
these, a subset of 8958 sources with 13 observed lenses forms a
well-defined subsample suitable for statistical analysis (Browne et
al.\ 2003, Table 13).  In particular, the CLASS statistical subsample
finds no lensing events with angular separations $\theta> 3''$.  An
explicit search at $6''< \theta < 15''$ also finds no lensing events
(Phillips et al.\ 2002).

In this paper we study the constraints from strong lensing statistics
in the post-WMAP era.  We examine two of the currently most uncertain
cosmological parameters: the equation of state of dark energy $w$ and
the power spectrum normalization $\sigma_8$, as well as two of the
most uncertain parameters in current lens models: the inner mass
density profile of clusters $\beta$ and the mass scale $M_c$ that
separates galaxy from cluster lenses.  $N$-body simulations indicate
that the dark matter profile in clusters has a roughly universal form
with an inner logarithmic slope of $\beta\sim 1$ to 1.5 (Navarro,
Frenk \& White 1997; Moore et al 1999).  However, recent analyses
based on observations of arcs in clusters (e.g.\ Sand, Treu \& Ellis
2002) indicate a shallower inner profile.  This uncertainty has
motivated us to treat $\beta$ for clusters as a free parameter in this
study.  For the galaxy-scale lenses with lower masses ($M<M_c$), we
assume the SIS model in accordance with the results cited above.  This
two-population galaxy-cluster model is needed to account for the large
number of observed small-separation lenses and the lack of
$\theta>3''$ systems in JVAS/CLASS (Keeton 1998; Porciani \& Madau
2000).

For the cosmological parameters, we use the results from WMAP in
conjunction with ACBAR, CBI and 2dF experiments (Spergel et al.\
2003).  We use the mean values $n=0.97$, $\Omega_M h^2=0.134$ and
$\Omega_B h^2=0.023$ (each measured to better than 5\%), and
marginalize over $\Omega_M=0.27\pm 0.04$ and $\sigma_8=0.84\pm 0.04$,
although we find the strength of our final constraints on $\beta$ and
$M_c$ to be insensitive to the strength of the particular priors used
on $\Omega_M$ and $\sigma_8$.  We consider only flat cosmological
models.

\section{Lens Profiles and Optical Depth}

We write the total optical depth at a given lens mass as the sum of
the optical depths due to SIS and generalized NFW (GNFW) halos:
\begin{equation}
\tau(M) = f_{SIS}(M)\tau_{\sss SIS}(M) + (1-f_{SIS}(M))\tau_{\sss
GNFW}(M)
\label{eq:tau}
\end{equation}

\noindent where $f_{SIS}$ is the fraction of halos that are SIS, and
$1-f_{SIS}$ is the fraction that are GNFW.  In addition to these two
populations, there is some evidence that objects with $M\la 10^{11}
M_{\sun}$ have shallow profiles (Ma 2003, Li \& Ostriker 2003). Ma
(2003), for example, derives a form for $f_{SIS}$ by requiring the
halo mass function and galaxy luminosity function to give consistent
optical depth predictions at subsecond scale.  Current lensing surveys
are limited to $\theta\ga 0.3''$, which cannot constrain small-mass
halos.  We therefore use the simplest possible form: $f_{SIS}=1$ for
$M<M_c$ and 0 otherwise, and leave $M_c$ as a free parameter to be
constrained by data.

An SIS lens has a density profile $\rho(r)=\sigma_v^2/(2\pi G r^2)$,
where $\sigma_v$ is the 1-d velocity dispersion.  SIS lenses produce
an image separation of $2\theta_E$, where $\theta_E=4\pi
(\sigma_v/c)^2 D_{ls}/D_s$ is the Einstein radius, and the cross
section is $\sigma_{\rm lens}=\pi (\theta_E D_l)^2 = 16\pi^3
(\sigma_v/c)^4 (D_lD_{ls}/D_s)^2$, where $D_s, D_l$, and $D_{ls}$ are
the angular diameter distances to the source, to the lens, and between
the lens and the source, respectively.  Using $\sigma_v=(\pi G^3 M^2
\Delta_{vir}\rho_M/6)^{1/6}$, we can then relate the cross-section
$\sigma_{\rm lens}$ to the halo mass $M$ which is needed for
eq.~(\ref{eq:tau}).  The GNFW profile (Zhao 1996) is given by
  
\begin{equation}  
\rho(r) = {\rho_s \over \left (\ds{r /r_s}\right )^\beta    \left
[1+\left (\ds{r/ r_s}\right )\right ]^{3-\beta}}  \,.
\end{equation}   
  
\noindent 
The mass of a halo is defined to be the virial mass $M
=4\pi\Delta_{vir}\rho_M r^3_{vir}/3$, where $r_{vir}$ is the virial
radius within which the average density is $\Delta_{vir}\rho_M$ and
$\rho_M$ is the mean matter energy density in the universe.  We
compute $\Delta_{vir}$ from the spherical-collapse model using the
fitting formula in Weinberg \& Kamionkowski (2003).  In practice, we
replace the scale radius $r_s$ by a concentration parameter $c(z)
\equiv r_{vir}(z)/ r_{s}(z)$, which is well described (for $\beta=1$)
by $c(z)=c_0 (1+z)^{-1}(M/M_*)^{-0.13}$ with $c_0=9$, where $M_*$ is a
typical collapsing mass for that cosmology at redshift zero (e.g.\
Bullock et al.\ 2001).  For $\beta\neq 1$, we compute $c_0$ by
assuming that the ratio $r_{1/2}/r_{vir}$ is independent of $\beta$,
where $r_{1/2}$ is the half-mass radius defined as
$M(r<r_{1/2})=M(r<r_{vir})/2$ (Li and Ostriker 2002).  The same
redshift and mass dependence is used for all $\beta$.

The optical depth for lensing is given by (e.g., Turner, Ostriker \&
Gott 1984)

\begin{equation}
\tau = \int_0^{z_s} dz_l {dD_l\over dz_l}\, (1+z_l)^3 \int_0^\infty dM
 {dn\over dM}(M, z_l)\,\sigma_{\rm lens}(M, z_l) B
\end{equation}

\noindent where $n(M,z_l)$ is the physical number density of dark
halos at the lens redshift $z_l$, $\sigma_{\rm lens}(M,z_l)$ is the
lensing cross section of a halo of mass $M$ at $z_l$, and $B$ is the
magnification bias.  We considered accounting for the ellipticity of
lenses, but found that for the JVAS/CLASS luminosity function,
ellipticity makes small ($\lesssim 10\%$) correction to the optical
depth, and therefore we choose to ignore it. This is consistent with
previous findings that ellipticity mostly affects image multiplicities
(e.g.\ Kochanek \& Blandford 1987; Wallington \& Narayan 1993).

To model the number density of lenses, we use the halo mass function
instead of the galaxy luminosity function (for the latter approach to
constrain the dark energy equation of state, see Chae et al.\ 2002).
The former has the advantages that (1) it accounts for all objects --
dark and luminous -- in the universe; (2) it has been accurately
calibrated by $N$-body simulations; and (3) its main dependence is on
now well-measured cosmological parameters instead of the more
uncertain astrophysical parameters.  We use the fitting formula of
Jenkins et al.\ (2001) for $dn/dM$ but note that their halo mass is
independent of cosmology and is defined to be the mass enclosed in
radius $r_{vir}$ within which the average density is
($180\rho_{crit}$), instead of ($\Delta_{vir}\rho_M$) based on the
spherical collapse model for general cosmology discussed earlier.  We
use the latter to define our halo mass in eq.~(3), but for
consistency, we convert it to $M_{180}$ and evaluate the fitting
formula for $dn/dM$ at $M_{180}$.  We find this step to lead to a
relatively small upward correction to the optical depth.  The matter
power spectrum is an input to $dn/dM$; we use the formulae of
Eisenstein \& Hu (1997) for $w=-1$ models and those of Ma et
al. (1999) for models with $w>-1$.  We do not consider $w<-1$ since
the power spectrum for these models is uncertain.

We include a scatter of $0.14$ in $\log_{10}c(z)$ (Bullock et
al. 2001; note that the published version of this paper incorrectly
quotes 0.18 for the scatter) for the halo concentration in eq.~(3).
We find that including this effect is very important, as it increases
the GNFW optical depth by more than an order of magnitude, in rough
agreement with Chen (2003) and Kuhlen, Keeton \& Madau (2003).

To compute the magnification bias, we note that the sources in CLASS
are well represented by a power-law luminosity function, $\phi(S) =
dn/dS \propto S^{-\eta}$, with $\eta \simeq 2.1$, which leads to $B =
4.76$ for all SIS lenses (Rusin \& Tegmark 2001).  
The magnification bias is more complicated for GNFW lenses.  We use
the formula in Oguri et al.\ (2002), which agrees very well with
ray-tracing simulations.

\begin{figure}[!t]
\includegraphics[height=3.7in, width=2.8in, angle=-90]{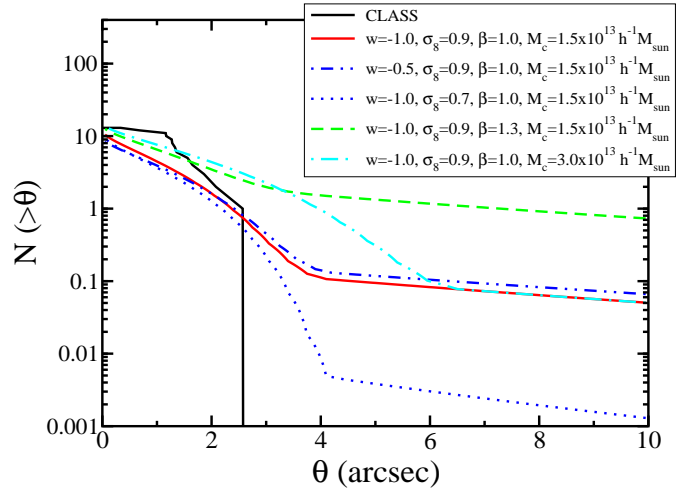}  
\caption{Number of lenses with angular separation above $\theta$ from
the CLASS statistical sample (Browne et al.\ 2003) and predictions
from the compound lens model with different values of $w$, $\sigma_8$,
$\beta$ and $M_c$.  Halos with $M<M_c$ are assumed to have the SIS
profile.  
}
\label{fig:P_theta}  
\end{figure}  

\section{Results}
  
Before discussing our statistical constraints on $\beta$, $M_c$,
$\sigma_8$, and $w$, we first illustrate in Fig.~\ref{fig:P_theta} the
dependence of lensing probability on these parameters by comparing the
expected number of lenses $N(>\theta)$ in various models with the
CLASS result.  It shows that the value of $M_c$ and an upper limit on
$\beta$ can be determined very accurately because the CLASS histogram
has a sharp cut-off in $\theta$.  Variations in $w$, on the other
hand, have a fairly small effect at all scales.  Models with low
$\sigma_8$ underpredict the optical depth at $\theta \lesssim 3''$,
while those with both high $\sigma_8$ and $\beta$ overpredict the
optical depth for $\gtrsim 3''$. Finally, we find that $N(>\theta)$ is
insensitive to the value of $\Omega_M$.

For the statistical tests, we use information from both the total
lensing optical depth $\tau$ and the image separation distribution
$d\tau/d\theta$ from JVAS/CLASS.  We do not use the lens redshift
distribution due to selection effects that are presumed to be
significant in this test.  This approach is similar to that described
in Davis, Huterer \& Krauss (2002).  The combined likelihood for the
two tests is
 
\begin{equation}  
{\mathcal{L}} = {N^x \exp(-N)\over x!}\times \prod_{i=1}^{x} \,
	 {1\over \tau_i}\,  
	\left.{d\tau\over d\theta} \right|_{\theta_i}     
\end{equation}  

\noindent where the first term accounts for the total optical depth
and the second for the angular distribution. Here $x=13$ is the number
of lenses in CLASS, $N=8958\tau$ is the number of galaxies predicted
by the model, and $\tau(w, \sigma_8, \beta, M_c, z_s)$ is the
predicted optical depth computed from eqs.~(1) and (3).  $\tau_i$ is
the optical depth {\it given} the source redshift $z_s$ for the lens
in question. When $z_s$ for a particular lens is not available, we set
$\tau_i=\tau$, i.e., we set $z_s$ to equal the mean redshift of the
whole source population, $\langle z_s\rangle=1.27$ (Marlow et al.\
2000).

\begin{figure}[!t]
\includegraphics[height=3.7in, width=3in, angle=-90]{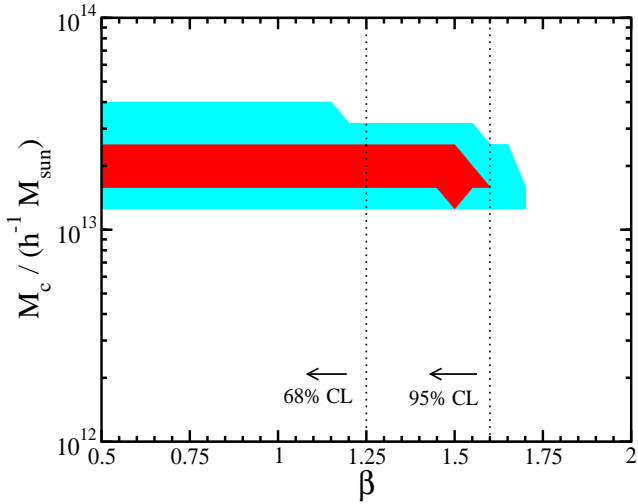}
\caption{Shaded regions show the 68\% and 95\% CL
likelihood joint constraint on the slope $\beta$ of the density
profile of $M>M_c$ halos and the mass $M_c$ separating the SIS and
GNFW lenses.  The results have been marginalized over all other
parameters using WMAP priors.  The vertical lines show the upper limits
on $\beta$ after further marginalization over $M_c$.}
\label{fig:beta_Mc}
\end{figure}

Fig.~\ref{fig:beta_Mc} shows the joint constraints on $\beta$ and
$M_c$ using the WMAP priors on $\Omega_M$ and $\sigma_8$ and the WMAP
mean values for $n$, $\Omega_M h^2$ and $\Omega_B h^2$ that are
determined very accurately.  Although we have fixed $w=-1$ here, we
have checked that the constraints are insensitive to $w$ by repeating
the analysis for several values of $w$ and finding essentially
identical results.  The absence of $\theta \gtrsim 3''$ lenses in
CLASS places an upper limit on the optical depth at these angular
scales, which in turn limits how steep the inner cluster density
profile can be: we find $\beta<1.25\,(1.60)$ at 68\% (95\%) CL (vertical
lines) after marginalization over $M_c$.  The constraints on $M_c$ are
also strong due to the sudden break in the predicted optical depth at
the angular separation corresponding to $M_c$ (see Fig.~1): we find
$1.58 < M_c/(10^{13}\, h^{-1}M_{\sun}) < 2.51$ at 68\% CL and $1.25 <
M_c/(10^{13}\, h^{-1}M_{\sun}) < 3.98$ at 95\% CL after
marginalization over $\beta$.

\begin{figure}[!t]
\includegraphics[height=3.7in, width=3in, angle=-90]{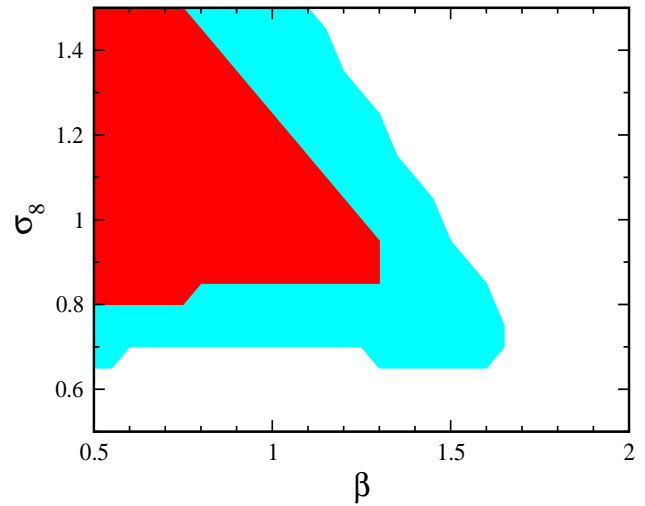}
\caption{68\% and 95\% CL likelihood joint constraints on $\beta$ and
$\sigma_8$ after marginalization over $M_c$ and $\Omega_M$. 
}
\label{fig:beta_sig8}
\end{figure}  

Fig.~\ref{fig:beta_sig8} shows the joint constraints on $\beta$ and
$\sigma_8$ using the same WMAP results as in Fig.~\ref{fig:beta_Mc}
and after marginalization over $M_c$.  It shows that steep density
profiles ($\beta \sim 1.5$) are consistent with the absence of
large-separation lenses only for a small range of low $\sigma_8$.
High $\sigma_8$ ($\gtrsim 1$), on the other hand, is allowed by CLASS
data provided the density profile is shallow.  The lower limit of
$\sigma_8>0.7$ at $95\%$ CL is insensitive to the value of $\beta$
because lowering $\sigma_8$ further would severely underpredict the
cumulative optical depth at small image separations.

We have also attempted to obtain joint constraints on $\sigma_8$ and
$w$ using the same WMAP parameters.  After marginalizing over
$\sigma_8$, our likelihood analysis gives $w<-0.75 (-0.55)$ at 68\%
(95\%) CL.  The solid and double-dotted curves in Fig.~1, however,
show that for two models with $w=-1$ and $-0.5$ but otherwise
identical parameters, we would expect nearly identical number of
lenses at $\theta< 2''$ and at most a factor of two of difference at
larger $\theta$.  This insensitivity to $w$ is due to the fact that
the matter power spectrum for different $w$ has identical shapes
except on very large length scales (see, e.g., Ma et al. 1999), so
once the models are normalized to the same $\sigma_8$ today, the only
difference is in the growth rates at higher redshift and in
$\Delta_{vir}$ for halo masses.  (See Sarbu, Rusin \& Ma 2001 for
constraints on $w$ for COBE-normalized power spectrum.)  The reason we
were able to place {\it any} constraints on $w$ is because the
theoretically computed optical depth for a fiducial (WMAP-favored)
cosmology {\it underpredicts} the total optical depth, so that various
models lie on the tail of the likelihood function.  Since a more
negative $w$ leads to a higher $\tau$, the total optical depth part of
the likelihood function exponentially suppresses models with $w>-1$.
We caution that constraints on $w$ (or any other parameters to which
the test in question is only weakly sensitive) are robust only when
the theoretically computed optical depth roughly agrees with the
observation.

\section{Discussion}

We have used the statistics of the recently completed JVAS/CLASS radio
survey to constrain the parameters in the two-population model of
halos in the universe. Motivated by recent evidence from direct
observations, $N$-body simulations, and semi-analytic arguments, we
have assumed that all objects with mass less than $M_c$ have the SIS
profile, while the more massive ones have the GNFW profile that scales
as $r^{-\beta}$ in the inner region of the halos. The absence of
$\theta \gtrsim 3''$ lenses in CLASS enables us to obtain tight
constraints on $M_c$ and an upper limit $\beta< 1.6$ at 95\% CL.
Furthermore, we obtain a constraint of $\sigma_8>0.7$ (95\% CL) on the
power spectrum normalization.

The constraints in our study have come from three effects: the total
lensing optical depth measured by CLASS, the shape of the image
separation distribution, and the lack of $\theta > 3''$ lenses.  It is
therefore interesting to consider what happens when at least one
large-separation event ($\gtrsim 5''$) is observed, and {\it when}
such an event will be observed given our knowledge of the halo density
profiles and cosmology. Keeton \& Madau (2001) ask a similar question,
and remark that one such large-separation event should be found in
surveys such as the Two-Degree Field and Sloan Digital Sky Survey
(SDSS), or else the cold dark matter model would have to be
questioned. Fig.~\ref{fig:P_theta}, however, shows that one could
easily accommodate an extremely small probability of finding a
large-separation event if, for example, we lived in a universe with
$\sigma_8\sim 0.7$.

This issue may soon be resolved: there is evidence for a
$\theta\approx 13''$ quadruple lens system in the SDSS (N. Inada et
al., in preparation).  Since this event is not yet a part of a
controlled survey with significant statistics, it cannot be included
in our analysis.  To test how the constraint on $\beta$ would change
with a positive detection of wide-separation lenses, we repeat our
analysis by assuming a 14th lens in the statistical sample of
JVAS/CLASS with $\theta=13''$ and a source at $z_s=2$.  We find this
hypothetical sample to favor $1.05<\beta < 1.35$ at 68\% CL, while
tightening the constraint on $M_c$ with its central value unchanged.
This result illustrates that stronger (lower as well as upper) limits
on $\beta$ would be obtained if we had at least one large angular
separation event in a controlled survey.  This is simply because
models with $\beta\lesssim 1$ predict $\ll 1$ lenses at large $\theta$
and would be strongly disfavored.

These factors motivate us to consider future surveys that may have
enough statistics to detect large-separation lenses.  The proposed LOw
Frequency ARray (LOFAR; www.lofar.org) is expected to find millions of
radio sources to fluxes below $1\mu Jy$ (Jackson 2002).  Although the
proposed angular resolution of $\sim 1''$ is inferior to that of
CLASS, LOFAR would be a useful survey for detecting large-separation
systems.  Estimates from Jackson (2002) indicate that LOFAR would have
17,000 lenses in the brighter part of the survey with $30\sigma$
signal-to-noise ratio, providing a thousandfold increase in lensing
statistics (if the identity of the lens events can be confirmed).
Given the small number of lenses in current surveys, even if the LOFAR
estimates are overly optimistic we can still expect that significantly
larger statistical lens samples and other future wide-field telescopes
such as SNAP (snap.lbl.gov) and LSST
(www.dmtelescope.org/dark\_home.html) will greatly improve our
understanding of the populations of objects in the universe.

We thank Mike Kuhlen and Chuck Keeton for numerous useful suggestions
and communicating their results prior to publication.  We thank Josh
Frieman for useful discussions.  DH is supported by the DOE grant to
CWRU. C-P Ma is supported by NASA grant NAG5-12173, a Cottrell
Scholars Award from the Research Corporation, and an Alfred P. Sloan
fellowship.  A portion of this work was carried out at the Kavli
Institute for Theoretical Physics.


\begin{thebibliography}{99}  

\bibitem{Bullock}  
	Bullock, J. S. et al.\ 2001, ApJ, 550, 21  

\bibitem{CLASS2}
	Browne, I.W.A. et al.\ 2003, MNRAS, 341, 13

\bibitem{chae}
	Chae, K.-H. 2003, MNRAS, MNRAS, 346, 746
  
\bibitem{chae2}
	Chae, K.-H. et al.\ 2002, Phys. Rev. Lett., 89, 151301
  
\bibitem{chen}
	Chen, D.-M. 2003, ApJ, 587, L55 

\bibitem{cheng_krauss_1}  
	Cheng, Y.-C.\ N.\ \& Krauss, L.\ M.\ 2000, 
	Int. J. Mod. Phys. A., 15, 697 
  
\bibitem{cohn}   
	Cohn, J.D., Kochanek, C.S., McLeod, B.A., \& Keeton, C.R. 2001,   
	ApJ, 554, 1216
    
\bibitem{Davis}  
	Davis, A.\ N., Huterer, D. \& Krauss, L.\ M.\ 2002, MNRAS, 
	344, 1029

\bibitem{eis}
	Eisenstein, D. \& Hu, W. 1997, ApJ, 511, 5  

\bibitem{LOFARmemo}  
	Jackson, N., ``LOFAR and Gravitational Lenses'' memorandum 
  
\bibitem[Jenkins et al. 2001]{VIRGO}
	Jenkins, A.\ R.\ et al.\ 2001, MNRAS, 321, 372

\bibitem{keeton}  
	Keeton, C.\ R., 1998, PhD. thesis, Harvard University

\bibitem{}  
	Keeton, C.R. \& Madau, P. 2001, ApJ, 549, L25  

\bibitem{king.jvas}  
	King, L.J., Browne, I.W.A., Marlow, D.R., Patnaik, A.R., \&  
	Wilkinson, P.N. 1999, MNRAS, 307, 225  
  
\bibitem{koch96}  
	Kochanek, C. S. 1996, ApJ, 466, 638  

\bibitem{koch87}  
	Kochanek, C. S. \& Blandford, R.  1987, ApJ, 321, 676
    
\bibitem{kkm}
  	Kuhlen, M., Keeton,  C.R. \& Madau, P. 2003, ApJ, in press

\bibitem{li_ostriker}  
	Li, L.-X. \& Ostriker, J.\ P.\ 2002, ApJ, 566, 652  

\bibitem{li_ostriker2}  
	Li, L.-X. \& Ostriker, J.\ P.\ 2003, ApJ, 595, 603

\bibitem{sch_sch}  
	Ma, C.-P. 2003, ApJ, 584, L1 

\bibitem{Ma_QCDM}  
	Ma, C.-P., Caldwell, R.R., Bode, P.\ \& Wang, L.\ 1999, 
	ApJ, 521, L1 

\bibitem{redshifts}  
	Marlow, D.R. et al.  2000, AJ, 119, 2629  
  
\bibitem{moore99}
	Moore, B., Quinn, T.,  Governato, F., Stadel, J. \& Lake, G. 1999,
	MNRAS, 310, 1147		
  
\bibitem{CLASS1}   
	Myers, S.T., et al.\ 2003, MNRAS, 341, 1

\bibitem{NFW97}
	Navarro, J.F., Frenk, C.S. \& White, S.D.M. 1997, ApJ, 490, 493 

\bibitem{oguri_arcs}
	Oguri, M., Taruya, A. \& Suto, Y. 2001
	ApJ, 559, 572

\bibitem{oguri_bias}
	Oguri, M., Taruya, A., Suto, Y. \& Turner, E.L. 2002, 
	ApJ, 568, 488

\bibitem{6-15arcsec}  
	Phillips, P.M., et al.\ 2001, MNRAS, 328,1001  

\bibitem{porciani}
	Porciani, C. \& Madau, P. 2000, ApJ, 532, 679

\bibitem{rusin_koch_keet}  
	Rusin, D., Kochanek, C.S.\ \& Keeton, C.R.\ 2003, ApJ, 
	595, 29

\bibitem{rusin_ma}  
	Rusin, D., \& Ma, C.-P. 2001, ApJ, 549L, 33  
    
\bibitem{rusin_tegmark}  
	Rusin, D. \& Tegmark, M.\ 2001, ApJ, 553, 709  

\bibitem{arcs}
	Sand, D., Treu, T., \& Ellis, R.  2002, ApJ, 574, L129

\bibitem{Sarbu}
	Sarbu, N., Rusin, D., \& Ma, C.-P. 2001, ApJ, 561, L147  

\bibitem{WMAP}
	Spergel, D.N. et al.\ 2003, ApJ, 148S, 175

\bibitem{takahashi}  
	Takahashi, R.\ \& Chiba, T.\ 2001, ApJ, 563, 489  
  
\bibitem{treu}  
	Treu, T.\ \& Koopmans, L.V.E.\ 2002, ApJ, 575, 87  

\bibitem{T90}  
	Turner, E. L.  1990, ApJ, 365, L43

\bibitem{TOG}  
	Turner, E. L., Ostriker, J.P., \& Gott, J.R. 1984, ApJ, 284, 1  
  
\bibitem{wall}
	Wallington, S. \& Narayan, R. 1993, ApJ, 403, 517

\bibitem{wk}
	Weinberg, N. \& Kamionkowski, M. 2003, MNRAS, 341, 251

\bibitem{winn}  
	Winn, J., Rusin, J. \& Kochanek C.S. 2003,  ApJ, 587, 80
  
\bibitem{Zhao}  
	Zhao, H.\ S.\ 1996, MNRAS, 278, 488  
  
\end{thebibliography}
\end{document}